\documentclass[twocolumn,showpacs,preprintnumbers,amsmath,amssymb]{revtex4}

\usepackage{epsfig}
\usepackage{graphics}
\usepackage{graphicx}
\usepackage{dcolumn}
\usepackage{bm}

\begin{document}

\preprint{APS/123-QED}

\title{Phase synchronization in time-delay systems}

\author{D.~V.~Senthilkumar$^1$}
\author{M.~Lakshmanan$^1$}%
 \email{lakshman@cnld.bdu.ac.in}
 \author{J.~Kurths$^2$}
 \email{jkurths@gmx.de}
\affiliation{%
$^1$Centre for Nonlinear Dynamics,Department of Physics,
Bharathidasan University, Tiruchirapalli - 620 024, India\\
}%
\affiliation{%
$^2$ Institute of Physics, University of Potsdam, 
Am Neuen Palais 10, 14469 Potsdam, Germany\\
}%

\date{\today}

\begin{abstract} 
Though the notion of phase synchronization has been well studied in
chaotic dynamical systems without delay, it has not been realized yet in
chaotic time-delay systems exhibiting non-phase coherent hyperchaotic
attractors.  In this article we report the first identification of phase
synchronization in coupled time-delay systems exhibiting hyperchaotic
attractor.  We show that there is a transition from non-synchronized behavior
to phase and then  to generalized synchronization as a function of coupling
strength. These transitions are characterized by  recurrence quantification
analysis, by phase differences based on a new transformation of the attractors
and also by the changes in the Lyapunov exponents. We have  found these
transitions in coupled piece-wise linear and in Mackey-Glass time-delay
systems. 
\end{abstract}

\maketitle


Synchronization is a natural phenomenon that one encounters in daily life. 
Since the identification of chaotic synchronization \cite{hfty1983,hfty198370,
lmptlc1990}, several papers have appeared in identifying and
demonstrating basic kinds of synchronization both theoretically and
experimentally ~(cf.\cite{asp2001,sbjk2002}). Among them, chaotic phase
synchronization (CPS) refers to the coincidence of characteristic time scales
of the coupled systems, while their amplitudes of oscillations remain chaotic
and often uncorrelated.  Phase synchronization (PS) plays a crucial role in
understanding the behavior of a large class of weakly interacting dynamical
systems in diverse natural systems. Examples include circadian rhythm,
cardio-respiratory systems, neural oscillators, population dynamics, electrical
circuits, etc  ~\cite{asp2001,sbjk2002,jk2000}.

The notion of CPS has been investigated so far in oscillators driven by
external periodic force \cite{aspgo1997,aspmgr1997}, chaotic oscillators with
different natural frequencies and/or with parameter mismatches
\cite{mgrasp1996,mgrasp1997,uplj1996,mzgww2002}, arrays of coupled chaotic
oscillators \cite{goasp1997,mzzz2000} and  also in essentially different
chaotic systems \cite{ercmt2003,sgchl2005}. On the other hand PS in nonlinear
time-delay systems, which form an important class of dynamical systems, have
not yet been identified  and addressed. A main problem here is to define 
even the notion of  phase in time-delay systems due to the intrinsic
multiple characteristic time scales in these systems.  Studying PS in
such chaotic  time-delay systems is  of considerable importance in many
fields,  as  in understanding the behavior of nerve cells (neuroscience), where
memory effects play a  prominent role,  in pathological and physiological
studies, in ecology, in lasers  etc
~\cite{asp2001,sbjk2002,jk2000,mcmlg1977,thif2003,nkgbe2000,mkph2005,lbsibs2006}.

In this paper, we report the first identification of phase synchronization (PS)
in nonidentical time-delay systems in the hyperchaotic regime with
non-phase coherent attractors with unidirectional nonlinear coupling.  We
will show the entrainment of phases of a coupled piecewise linear time-delay
system for weak coupling from the non-synchronized state. Phase is calculated
using the Poincar\'e method \cite{asp2001,sbjk2002}  after a new transformation
of attractors of the time-delay systems, which looks  then like a smeared limit
cycle. The existence of PS and generalized synchronization (GS) in coupled
time-delay systems is characterized by recently proposed methods based on
recurrence quantification analysis and also in terms of Lyapunov exponents of
the coupled time-delay systems.  

\begin{figure}
\centering
\includegraphics[width=1.0\columnwidth]{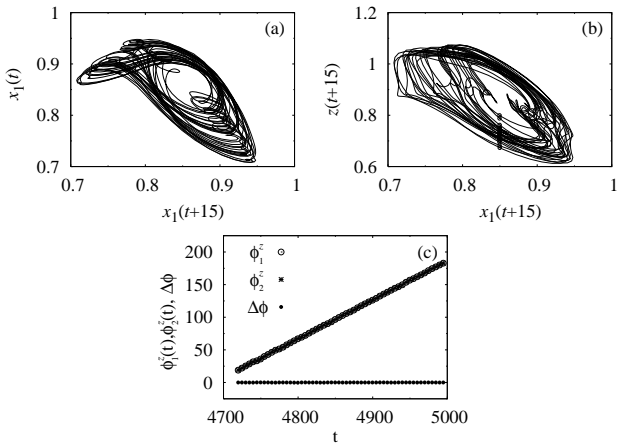}
\caption{\label{chaos} Phase synchronization between the systems
(\ref{eq.one}a) and (\ref{eq.one}b). (a) The non-phase coherent hyperchaotic
attractor of the uncoupled drive (\ref{eq.one}a). (b) Transformed attractor in
the $x_1(t+\tau)$ and $z(t+\tau)$ space along  with the Poincar\'e points
represented as open circles. (c) The phases of the drive system
($\phi_1^{z}$), the response system ($\phi_2^{z}$) calculated from
the new state variable $z(t+\tau)$ and their difference ($\Delta \phi$) for
$b_3=1.5$.}
\end{figure}

We first consider the following unidirectionally coupled drive $x_1(t)$ and response
$x_2(t)$ systems, which we have recently studied in detail in
\cite{dvskml2005,dvskml2005jp,dvskml2005ijbc},
\begin{subequations}
\begin{eqnarray}
\dot{x_1}(t)&=&-ax_1(t)+b_{1}f(x_1(t-\tau)),  \\
\dot{x_2}(t)&=&-ax_2(t)+b_{2}f(x_2(t-\tau))+b_{3}f(x_1(t-\tau)),
\end{eqnarray}
\label{eq.one}
\end{subequations}
where $b_1, b_2$ and $b_3$ are constants, $a>0$, $\tau$ is the delay time and 
$f(x)$ is the piece-wise linear equation of the form
\begin{eqnarray}
f(x)=
\left\{
\begin{array}{cc}
0,&  x \leq -4/3  \\
            -1.5x-2,&  -4/3 < x \leq -0.8 \\
            x,&    -0.8 < x \leq 0.8 \\              
            -1.5x+2,&   0.8 < x \leq 4/3 \\
            0,&  x > 4/3. \\ 
         \end{array} \right.
\label{eqoneb}
\end{eqnarray}
We have chosen the values of parameters as $a=1.0,b_1=1.2,b_2=1.1$ and
$\tau=15$, which are outside the region of complete, lag and anticipatory
synchronizations discussed in  \cite{dvskml2005,dvskml2005jp}. For this
parametric choice, in the absence of coupling,  the drive $x_1(t)$ and the
response $x_2(t)$ systems evolve independently. Further in this case, the drive
$x_1(t)$  exhibits a hyperchaotic  attractor (Fig.~\ref{chaos}a) with five
positive Lyapunov exponents (See \cite{dvskml2005} for the spectrum of Lyapunov
exponents) and the response $x_2(t)$ has four positive Lyapunov exponents, i.e.
both subsystems are qualitatively different  (due to $b_1\ne b_2$).  The
parameter $b_3$ is the coupling strength of the  unidirectional nonlinear
coupling (\ref{eq.one}b), while the parameters $b_1$ and $b_2$ play the role of
parameter mismatch resulting in nonidentical coupled time-delay systems.   

Now the important questions we encounter are whether PS exists in the
time-delay system (\ref{eq.one}) when the coupling is included ($b_3>0$) and,
if so, how to characterize the possible transition to PS in such systems which
possess in general highly non-phase  coherent attractors having a broad-band
power spectrum.  In low-dimensional systems, a few methods
\cite{asp2001,sbjk2002} have been developed  to define phase in  phase coherent
chaotic attractors, which have a dominant peak in the power spectrum.
Definition of phase is not so clear in non-coherent chaotic attractors, in
particular in high-dimensional systems having broad-band power spectra such as
time-delay systems.  Methods to calculate  phase of non-coherent attractors of 
time-delay systems is not readily available.   One approach to calculate the
phase of system  (\ref{eq.one}) is based on the concept of curvature
\cite{gvobh2003}, which is often used in low-dimensional systems.  However, we
find that this procedure does not work in the case of  time-delay systems   in
general, and in particular for Eqs.~(\ref{eq.one}). We present here three other
approaches to study PS  in systems as (\ref{eq.one}):

i) We introduce a new transformation to successfully capture the phase in the
present problem.    It transforms the non-phase coherent attractor
(Fig.~\ref{chaos}a) into  a smeared limit cycle-like form with well defined
rotations around one center (Fig.~\ref{chaos}b). This transformation is
performed by introducing the new state variable 
\begin{align}
z(t+\tau)=x_1(t)x_1(t+\hat{\tau})/x_1(t+\tau),
\end{align}
where $\hat{\tau}$ is the optimal value of delay time to be chosen (so
as to rescale the original non-phase coherent attractor into a smeared limit
cycle-like form), we plot the above attractor (Fig.~\ref{chaos}a) in the 
($x_1(t+\tau),z(t+\tau))$ phase space.  The functional form of this
transformation has been identified by generalizing the transformation used in
the case of chaotic atractors in the Lorenz system \cite{asp2001}. We find the
optimal value of $\hat{\tau}$ to be $1.65$.  The above transformation is
obtained through a suitable functional form (along with a delay time
$\hat{\tau}$), so as to unfold the original attractor  (Fig.~\ref{chaos}a) into
a phase coherent attractor. Now the attractor (Fig.~\ref{chaos}b) looks indeed
like a smeared limit cycle with nearly well defined rotations around a fixed
center. Hence, we can calculate the phase using the Poincar\'e method
\cite{asp2001,sbjk2002} from the rescaled attractor. The Poincar\'e points are
shown as open circles in Fig.~\ref{chaos}b.  The phases of both the drive
$\phi_1^{z}(t)$ and the response $\phi_2^{z}(t)$ systems calculated from
the new state variable $z(t+\tau)$ are shown in Fig.~\ref{chaos}c along with
their phase difference $\Delta\phi=\phi_1^{z}(t)-\phi_2^{z}(t)$ for the value
of the coupling strength $b_3=1.5$, showing a high quality PS. This strong
boundedness of the phase difference obtains for $b_3 \ge 1.382$. (Note
that the transformed attractor (Fig.~\ref{chaos}b) does not  have any closed
loops as in the case of the original attractor (Fig.~\ref{chaos}a).  If it is
so, such closed loops will lead to phase mismatch, and one cannot obtain exact
matching of phases of both the drive and response systems as shown in
Fig.~\ref{chaos}c).

ii) Next, we analyze the complex synchronization phenomena in the coupled
time-delay systems (\ref{eq.one})  by means of very recently proposed methods
based on  recurrence plots \cite{mcrmt2005}. These methods help to identify and
quantify PS  (particularly in non-phase coherent attractors)  and GS. 
For this purpose, the generalized autocorrelation
function $P(t)$ \cite{mcrmt2005} was introduced
\begin{align}
P(t)=\frac{1}{N-t} \sum_{i=1}^{N-t} \Theta(\epsilon-||X_i-X_{i+t}|| ), 
\label{pbt}  
\end{align}
where $\Theta$ is the Heaviside function, $X_i$ is the $i$th data  of the
system $X$, $\epsilon$ is a predefined threshold. $||.||$ is the  Euclidean norm
and $N$ is the number of data points. Looking at the coincidence of the
positions of the maxima of $P(t)$ for both  systems, one can qualitatively
identify PS.

\begin{figure}
\centering
\includegraphics[width=0.84\columnwidth]{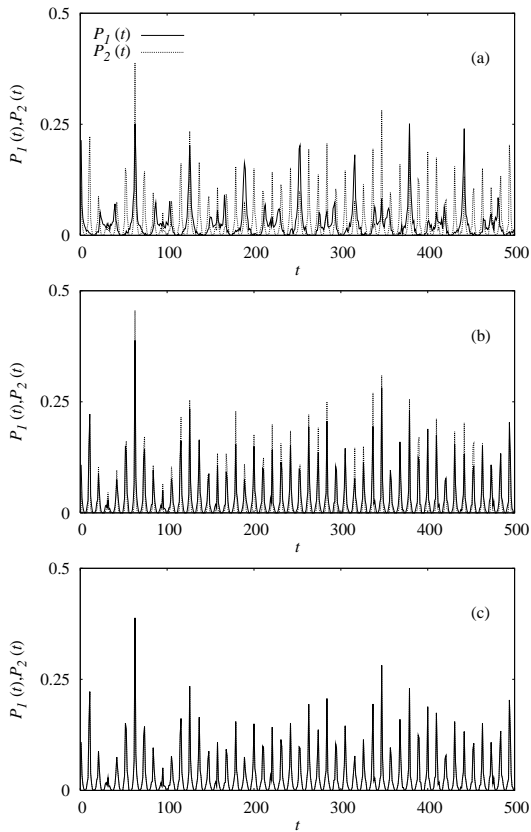}
\caption{\label{fig2} Generalized autocorrelation functions of both the drive
$P_1(t)$ and the response $P_2(t)$ systems. (a) Non-phase synchronization for
$b_3=0.6$, (b) Phase synchronization for $b_3=1.5$ and (c) Generalized
synchronization for $b_3=2.3$.}
\end{figure}

A criterion to quantify PS is the cross correlation 
coefficient between the drive, $P_1(t)$, and the response,
$P_2(t)$, which can be defined as Correlation of Probability of Recurrence
(CPR)
\begin{align}
CPR=\langle \bar{P_1}(t)\bar{P_2}(t)\rangle/\sigma_1\sigma_2,
\end{align}
where $\bar{P}_{1,2}$ means that the mean value has been subtracted and
$\sigma_{1,2}$ are the standard deviations of $P_1(t)$ and $P_2(t)$
respectively.  CPR $\approx 1$  indicates that the systems are in complete PS,
whereas for non-PS  one obtains low values of CPR.

To characterize GS, the authors of \cite{mcrmt2005} proposed the first index 
as the Joint Probability of Recurrences (JPR),
\begin{align}
JPR=\frac{\frac{1}{N^2} \sum_{i,j}^N \Theta(\epsilon_x-||X_i-X_j||
)\Theta(\epsilon_y-||Y_i-Y_j||)-RR}{1-RR}  
\label{jpr}  
\end{align}
where $RR$ is rate of recurrence, $\epsilon_x$ and $\epsilon_y$ are thresholds
corresponding to the drive and response systems respectively. RR measures
the density of recurrence points and it is fixed as 0.02 \cite{mcrmt2005}. JPR
is close to $1$ for systems in GS and is small when they are not  in GS. The
second index depends on the coincidence of probability of recurrence, which is
defined as Similarity of Probability of Recurrence (SPR),
\begin{align}
SPR=1-\langle(\bar{P_1}(t)-\bar{P_2}(t))^2\rangle/\sigma_1\sigma_2.
\end{align}  
SPR is of order $1$ if both the systems are in GS and approximately zero if
they evolve independently.

\begin{figure}
\centering
\includegraphics[width=1.0\columnwidth]{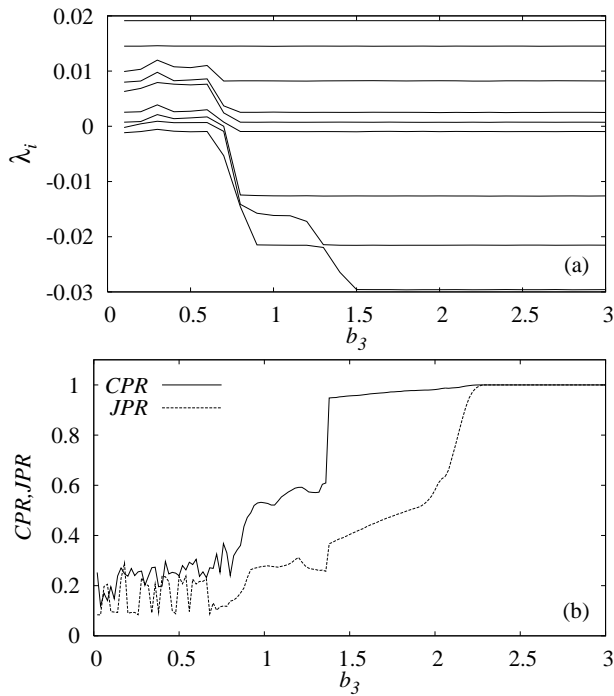}
\caption{\label{fig3} (a) Spectrum of first nine largest Lyapunov exponents  of
the coupled  systems (\ref{eq.one}) as a function of coupling strength $b_3 \in
(0,3)$  and (b) Spectrum of CPR and JPR as a function of coupling strength $b_3
\in (0,3)$.}
\end{figure}

Now, we will apply this concept to the original (non-transformed) attractor
(Fig.~\ref{chaos}a). We  estimate these recurrence based measures from $5000$
data points after sufficient transients with the integration step $h=0.01$ and
sampling rate $\Delta t =100$. The generalized autocorrelation functions
$P_1(t)$ and $P_2(t)$ (Fig.~\ref{fig2}a) for the coupling $b_3=0.6$ show that
the maxima of both systems do not occur simultaneously and there exists a
drift between them, so there is no synchronization at all.  This is also
reflected in the rather low value of CPR $=0.381$.  For $b_3 \in (0.91,1.381)$,  we
observe the first substantial increase of recurrence reaching CPR $\approx
0.5-0.6$. Looking to the details of  the generalized correlation functions 
$P(t)$, we find that now the main oscillatory dynamics becomes locked, i.e.
the main maxima of $P_1$ and $P_2$ coincide. For $b_3 \in (1.382,2.2)$ CPR
reaches almost $1$, i.e. now all maxima of $P_1$ and $P_2$ are in agreement
and this is in accordance with strongly bounded phase differences.  This is a
strong indication for PS.  Note, however that the heights of the peaks are
clearly different (Fig.~\ref{fig2}b).  The differences in the peak heights
exhibit that there is no strong interrelation in the amplitudes. Further
increase  of the coupling (here $b_3=2.3$) leads to the coincidence of both
the positions and the heights of the peaks (Fig.~\ref{fig2}c) referring to GS
in systems (\ref{eq.one}). This is also confirmed from the maximal values of
the indices JPR $=1$ and SPR $=1$, which is due to the strong correlation in
the amplitudes of both systems. The transition  from non-synchronized to PS
and then to GS is characterized by the maximal values of  CPR, SPR and JPR
(Fig.~\ref{fig3}b). As expected from the construction of these functions, CPR
refers mainly to the onset of PS, whereas JPR quantifies clearly the onset of
GS.  The existence of GS is also  confirmed using the auxiliary
system approach\cite{hdianfr1996}.

iii) The transition from non-synchronization to PS is also characterized by
changes in the Lyapunov exponents of the coupled time-delay systems
(\ref{eq.one}). The spectrum of the  nine largest Lyapunov exponents of the
coupled systems is shown in Fig.~\ref{fig3}a, from which one can find that the
Lyapunov exponents corresponding to the response system become negative from
the value of the coupling strength $b_3>0.9$, where the transition to PS
occurs, except for the largest Lyapunov exponent $\lambda_{max}^{(2)}$ which
continues to remain positive. This is a strong indication that in this rather
complex attractor the amplitudes become somewhat interrelated already at the
transition to PS (as in the funnel attractor \cite{gvobh2003}). It is
interesting to note that the Lyapunov exponents of the response system
$\lambda_i$ (other than $\lambda_{max}^{(2)}$) are changing already at the
early stage of PS ($b_3 \approx 0.91$), where  the  complete PS  is not yet
reached.  This has been also observed for the onset of PS in phase-coherent
oscillators ~\cite{mgrasp1996}.

We have obtained the same results for different sampling intervals $\Delta t$
and for various values of delay time $\tau$.  We have also identified this
transition to PS and to GS  in the coupled Mackey-Glass systems
\cite{mcmlg1977,jdf1982}, and also in (\ref{eq.one}) for linear coupling
(unidirectional) using these above three  approaches  (these results will  be
published in a more comprehensive paper).

In conclusion, we have identified for the first time the existence of PS in
coupled  time-delay systems in the hyperchaotic regime with highly
non-phase coherent attractors.  We have shown that there is a typical
transition from non-synchronized state to PS for weak coupling  and in the
range of strong coupling there is a transition to GS from PS. We have
also identified a suitable transformation, which works equally well for
Mackey-Glass system (having a more complex hyperchaotic attractor), to capture
the phase of the underling attractor. We have also characterized the existence
of PS and GS in terms of recurrence based indices like generalized
autocorrelation function $P(t)$, CPR, JPR and SPR and quantified the different
synchronization regimes in terms of them.  The above transition is also 
confirmed by the changes in the Lyapunov exponents. We have pointed out the
existence of PS in coupled Mackey-Glass systems as well.  The recurrence based
technique as well as the new transformation are also appropriate for the
analysis of experimental data, i.e. we expect experimental verification of
these findings.

\begin{acknowledgments}
The work of D. V. S and M. L has been supported by a
Department of Science and Technology, Government of India sponsored research
project. J. K has been supported by his Humboldt-CSIR research award and 
NoE BIOSIM
(EU).
\end{acknowledgments}


\end{document}